\documentclass[preprintnumbers, prd, onecolumn, showpacs,floatfix,preprintnumbers,
superscriptaddress, nofootinbib]{revtex4}
\usepackage{graphicx}
\usepackage{epsfig}
\usepackage{bm}
\usepackage{amssymb}
\usepackage{float}
\usepackage{amsmath}
\usepackage{dcolumn}
\usepackage{cancel}
\usepackage[colorlinks]{hyperref}
\usepackage[usenames,dvipsnames]{color}
\hypersetup{
     breaklinks=true,
    pdfstartview={FitH},    % fits the width of the page to the window
    colorlinks=true,       % false: boxed links; true: colored links
    linkcolor=blue,          % color of internal links
    citecolor=red,        % color of links to bibliography
    filecolor=magenta,      % color of file links
    urlcolor=blue,           % color of external links
    anchorcolor=green,      % Color for anchor text
    linktocpage=true
}

\def\doi{http://doi.org}

\begin{document}

\title{A Conceptual Model for the Origin of the Cutoff Parameter in Exotic Compact Objects}

\author{W. A. Rojas C}
\email{warojasc@unal.edu.co}
\affiliation{Departamento de F\'isica, Universidad Nacional de Colombia,  Bogot\'a, Colombia}

\author{J. R. Arenas S.}
\email{jrarenass@unal.edu.co}
\affiliation{Observatorio Astron\'omico Nacional\\ Universidad Nacional de Colombia, Bogot\'a, Colombia}

\newcommand{\be}{\begin{equation}}
\newcommand{\ee}{\end{equation}}
\newcommand{\bea}{\begin{eqnarray}}
\newcommand{\eea}{\end{eqnarray}}
\newcommand{\bc}{\begin{center}}
\newcommand{\ec}{\end{center}}
%%%%%%%%%%%%%%%%%%%%%%%%%%%%%%%%%%%%%%%%%%%%%%%%%%%%%%%%%%%%%%%%%%%%%%%%%%
%%%%%%%%%%%%%%%%%%%%%%%%%%%%%%%%%%%%%%%%%%%%%%%%%%%%%%%%%%%%%%%%%%%%%%%%%%
\begin{abstract}
A Black Hole (BH) is a spacetime region with a horizon and where geodesics converge to a singularity. At such a point, the gravitational field equations fail. As an alternative to the problem of the singularity arises the existence of Exotic Compact Objects (ECOs) that prevent the problem of the singularity through a transition phase of matter once it has crossed the horizon. ECOs are characterized by a closeness parameter or cutoff, $\epsilon$, which measures the degree of compactness of the object. This parameter is established as the difference between the radius of the ECO's surface and the gravitational radius. Thus, different values of $\epsilon$ correspond to different types of ECOs. If $\epsilon$ is very big, the ECO behaves more like a star than a black hole. On the contrary, if $\epsilon$ tends to a very small value, the ECO behaves like a black hole. It is considered a conceptual model of the origin of the cutoff for ECOs, when a dust shell contracts gravitationally from an initial position to near the Schwarzschild radius. This allowed us to find that the cutoff makes two types of contributions: a classical one governed by General Relativity and one of a quantum nature, if the ECO is very close to the horizon, when estimating that the maximum entropy is contained within the material that composes the shell. Such entropy coincides with the Bekenstein--Hawking entropy. The established cutoff corresponds to a dynamic quantity dependent on coordinate time that is measured by a Fiducial Observer (FIDO). Without knowing the details about quantum gravity, parameter $\epsilon$ is calculated, which, in general, allows distinguishing the ECOs from BHs. Specifically, a black shell (ECO) is undistinguishable from a BH
\end{abstract} 
%%%%%%%%%%%%%%%%%%%%%%%%%%%%%%%%%%%%%%%%%%%%%%%%%%%%%%%%%%%%%  
\maketitle 

%%%%%%%%%%%%%%%%%%%%%%%%%%%%%%%%%%%%%%%%%%%%%%%%%%%%%%%%%%
\section{Introduction}\label{Sec1}
A Black Hole (BH) corresponds to a solution of the gravitational field equations that describes a region where the spacetime curvature is so high that not even light can escape. There, the geodesic lines converge to a point referred to as a singularity. Specifically, a BH is characterized by a singularity surrounded by an event horizon, which hides the interior of the black holes from Fiducial Observers (FIDOs) \cite{Penrose:1969pc}. The fact that the geodesic lines meet at a point that is practically mathematical poses enormous difficulties, since it is not possible to physically describe what occurs there; the field equations are violated in the singularity, which is why the manifold is incomplete \cite{Penrose:1964wq}. As an alternative to the black hole paradigm arises the existence of Exotic Compact Objects (ECOs) \cite{Cardoso:2019rvt}, such as dark stars or gravastars. The latter are characterized by a de Sitter interior spacetime and a Schwarzschild exterior spacetime. On the border of the two regions, there is a thin shell of matter that coincides with the event horizon. In~this type of model, the gravitational collapse is subject to a transition phase that prevents future collapse~\cite{Cardoso:2019rvt,Chapline:2000en, S.:2016dul, Poisson:2009pwt}.

Experimental evidence of Gravitational Waves (GWs) has been found for compact objects \cite{Abbott:2016,Abbott:2020uma,Abbott:2017}, including the event GW190814, where the nature of a compact object is unknown and could be an exotic compact object \cite{Abbott:2020khf}. This has established a new field of observational astronomy, i.e., gravitational wave astronomy, which enables the study of binary systems, such as ECO-ECO, BH-BH, and ECO-BH.

Abedi et al. analyzed the gravitational waves GW150914, GW151226, and LVT151012%define if appropriate
; then, they suggested that GW signals arise due to merging BHs or ECO systems \cite{Abedi:2016hgu}. In addition, Cardoso et al. showed that in some configurations, the coalescence of ECOs (compact boson stars) might be almost indistinguishable from that of BHs \cite{Cardoso:2016oxy}. 

ECOs are of great importance, not only in research about the nature and existence of astrophysical BHs, but also the parameter $\epsilon$ that characterizes them would be associated with some basic scale of quantum gravity \cite{Addazi:2018uhd}. The observation of the coalescence of compact objects based on the study of gravitational radiation would aid in explaining the nature of quantum gravity. ECOs are inspired by the quantum effects of BHs and seek to provide an answer to BH problems, such as information loss and the singularity \cite{Barausse:2018vdb}.

An ECO can be characterized by the closeness parameter:
\begin{equation}
\epsilon=r-r_{Sch},
\end{equation}
for $r\sim r_{Sch}$, where $r$ describes the position of the surface of the ECO and $r_{Sch}=\frac{2GM}{c^2}$ is the radius associated with the event horizon \cite{Addazi:2018uhd,Cardoso:2019rvt}. This parameter allows quantifying how compact an ECO is \cite{Chen:2019hfg}.

$\epsilon$ is a quantity depending on the observer that is associated with the Love numbers \cite{Addazi:2018uhd}:
\begin{equation}
\epsilon=r_{Sch}e^{-1/k},
\end{equation}
where $k$ is one of the Love numbers. Hence, $\epsilon$ would allow distinguishing a BH from an ECO in general, but it would also show that a BH would be undistinguishable from a black shell.

In Newtonian physics, Love numbers relate the mass multipole moments created by tidal forces on a spherical body. These moments encode information on the body's internal structure, and it can be transported by the GWs. In other words, the deformity of an ECO induced by tidal forces is coded by Love numbers and sensitively depends on the ECO's internal structure. Determining Love numbers implies understanding the physics of the ECO \cite{Pani:2015hfa}.

In the case of an ECO, the Love numbers are different from zero and are encoded in the gravitational waves; in binary ECOs, they allow studying the behavior of matter with enormous densities, and for a static BH type as the Schwarzschild type, they are null, which leads to the multipole structure remaining undisturbed when immersed%please ensure that the original meaning is retained
 in a tidal field \cite{Binnington:2009bb,Pani:2015nua}. In 2017, Cardoso et al. determined that a Schwarzschild BH in Chern--Simons gravity has Love numbers different from zero \cite{Cardoso:2017cfl}. We expect that by characterizing the ECOs through Love numbers, the Love relationships contained in the GWs from the fusion of binary systems will allow understanding the behavior of the matter subjected to enormous densities and the physics of the GWs themselves \cite{Pani:2015nua,Yagi:2016ejg}. Recently, Le Tiec et al. estimated the Love numbers for a Kerr BH, which would imply that they would be susceptible to the deformity of an external tidal field and that such deformity could be detected in GW-LIGO \cite{LeTiec:2020spy}.

There are several conceptions of the parameter $\epsilon$. Among others, Guo et al. obtained, for one type of ECO referred to as a fuzzball, $\epsilon\ll GM$ close to $r_{Sch}$ and, for a firewall, $\epsilon \sim l^{2}_{p}/GM$ \cite{Guo:2017jmi}. It is interesting to mention the results with significant quantum effects such as \cite{Barausse:2018vdb}:
\begin{equation}
\epsilon =\frac{t_{P}}{t_{H}}l_{p},\,\,l_{p}=\sqrt{\frac{G\hbar}{c^{3}}}, \,\,t_{P}=\sqrt{\frac{\hbar G}{c^{5}}},\,\,t_{H}=\frac{GM}{c^3}.
\end{equation}

Furthermore, for an ECO (black shell), a model is introduced that associates the creation of particles and a closeness parameter \cite{Harada:2018zfg}:
\begin{equation}
\epsilon=\sqrt{1-\frac{2MG}{c^{2}r_{f}}}l_{p}\ll l_{p},
\end{equation}
where $r_{f}$ is the radius limit of approach.

The instability of the ECOs and their implications in the GWs were proposed by B. Chen et al., where an ECO has a radius:
\begin{equation}
r_{ECO}=rs+\epsilon,\,\,\,\Delta=\sqrt{\frac{8GM\epsilon}{c^{2}}}.
\end{equation}
where $\Delta$ is a supplementary characterization of the compactness of the ECO \cite{Chen:2019hfg}. Additionally, it~has been shown that ECOs are unstable due to the accretion of matter and to the influence of GWs on gravitational collapse, resulting in, when $\epsilon \sim l_{p}$, them necessarily converting into BHs. To prevent gravitational collapse, $\epsilon \gg l_{p}$ must be met, or the null conditions of energy are violated \cite{Addazi:2019bjz}. 

Black shells share the properties of radiation emission with the BHs when they are considered quantum fields in the vicinities of the black shells. Any shell of a mass $M$ that collapses until it is close to its gravitational radius emits thermal radiation, with a radius limit given by \cite{Paranjape:2009ib}:
\begin{equation}
r_{f}=\frac{2GM}{c^{2}\left(1-\frac{\epsilon^2}{l_{p}^{2}}\right)}.
\label{W99}
\end{equation}

In the case of a closeness parameter $\epsilon \longrightarrow 0$, a Boson Star (BS) possesses properties analogous to a BH, such as radiation at the Hawking temperature $T_{H}$ and a Bekenstein--Hawking entropy $S_{BH}$ \cite{Saravani:2012is, Barcelo:2010xk}. For other types of ECOS, stability differs; for example, gravastar ECOs are more thermodynamically stable in contrast to BHs \cite{Uchikata:2015yma}.

Within the wide spectrum of ECOs registered in detail in \cite{Cardoso:2019rvt}, which includes white dwarfs and neutron stars, to exotic objects such as quark stars, hybrid stars with gluon-plasma nuclei, superspinars, wormholes, etc., the Gravastars (GSs) and the Boson Stars (BSs) stand out, which as the black shells, are feasible alternatives to astrophysical BHs, but the GSs and BSs are unstable since they exhibit a very unstable ergoregion and have a very short life \cite{Cardoso:2008kj}.

Research on the nature of the ECOs goes beyond theoretical speculations because their study is being developed with models conceived of for observation. It is important to know the available observational methods to complete the compact object search scenario. A usual method to distinguish a neutron star from a BH is based on measuring their mass. In the case of an ECO, if its mass is greater than the Chandrasekhar limit, it is believed that the ECO is a BH. Nevertheless, luminosity criteria associated with mass measurements and the angular momentum are not entirely reliable methods given the broad spectrum of ECOs that could be found, which is why other techniques to distinguish them are needed. A promising technique to identify ECOs from BHs lies in the observation of GWs. It is thought that the study of GWs in the inspiral phase of the fusion of binary systems allows determining their mass and the multipole moments distinguishing an ECO from a BH \cite{Cardoso:2007az}. In 2016, Cardoso et al. studied the GW product of the fusion of two black shells of equal mass, and they compared them to the GWs corresponding to two BHs. They found that under certain configuration, the GW signals are almost undistinguishable \cite{Cardoso:2016oxy}. Another method to distinguish ECOs from BHs was proposed by Cardoso~et~al., where they considered that compact objects with rings of light are BHs, given that the light rings are associated with the photon sphere at a radius $r=\frac{3GM}{c^{2}}$, in contrast to the ECOs that do not exhibit these luminous rings \cite{Cardoso:2007az}. Current evidence leads to thinking that ECOs with high angular momentums are similar to a Kerr BH, but without the formation of event horizons \cite{Cardoso:2014sna}.

Based on this conceptualization of the ECOs in the observational context described, our main purpose in this paper is to contribute to the characterization of the nature of the ECOs in terms of the closeness parameter $\epsilon$. We propose a model that relates criteria from the quantum microscopic world of compact objects with observations corresponding to the astrophysical macroscopic world, as a basis for estimating $\epsilon$. 

With relativistic kinematic properties, referring to the observer's notion, we model a type of ECO referred to as a black shell, undistinguishable from a BH, and that is an alternative object to the BHs, as~would be observed by an FIDO.

$\epsilon$ was calculated based on quantum foundations that establish a connection with the notions of quantum gravity through the Bekenstein--Hawking entropy and that exhibit a semi-classic and another limit where the quantum effects are significant. 

$\epsilon$ is a measure of the compactness of the ECOs, which we expect to be an indicator to classify and distinguish compact objects in general, through the relationship existing between this parameter and the Love numbers. 

In Section \ref{Sec2}, the quantum origin of the closeness parameter $\epsilon$ that characterizes the ECOs is substantiated. Section \ref{Sec3}, describes that kinematic nature of the cutoff parameter for a black shell. The~quantum model of the closeness parameter is introduced in Section \ref{Sec4}, and Section \ref{Sec5} is devoted to summarizing and discussing this paper.

%------------------------------------------------------------------------------------------------------------------------------------------
\section{Quantum Mechanical Foundation of the Cutoff for a Black Shell}\label{Sec2}

Let there be a thin spherical dust shell that gravitationally contracts from a certain position $r_{0}$ to near its gravitational radius $r(t_{2})=r_{Sch}+\epsilon$, at a finite time $t_{2}$ measured by an FIDO observer.

The nature of the closeness parameter $\epsilon$ obeys a bound of a quantum nature and is closely related to the shell's entropy compacted close to the gravitational radius.

The origin of this entropy is understood in terms of the properties of the physical vacuum in strong gravitational fields. In the case of BHs, an observer that is at rest with respect to the horizon sees~the zero-point fluctuations of physical fields as a thermal atmosphere around a BH \cite{Fursaev:2004qz}. An~important property of the BHs is that the entanglement entropy $S_{ent}$ coincides with the entropy of that thermal atmosphere \cite{Mukohyama:1998rf,tHooft:1984kcu,Israel:1976, Zurek:1985,Fulling:1987}.

$S_{ent}$ may be a source of Bekenstein--Hawking entropy $S_{BH}$ for a species, if $\epsilon \neq 0$ in the expression~\cite{Mukohyama:1998rf,tHooft:1984kcu,Arenas:2011be}:
\begin{equation}
S_{ent} \sim \frac{A}{4\epsilon},
\label{W0}
\end{equation}
where $A$ is a surface very close to the event horizon, with its proper altitude $\eta$ related to the parameter $\epsilon$ through the relationship:
\begin{equation}
\epsilon=\frac{1}{2}\frac{\kappa_{0}}{c^{2}}\eta^{2}
\label{W1}
\end{equation}
and with $\eta^{2}\sim l^{2}_{p}$, $\kappa_{0}$ being the surface gravity and $l_{p}$ the Planck length.

Depending on $\epsilon$, without considering the dependency that $S_{BH}$ would have on all the fields present in nature, in principle, it can be identified with $S_{ent}$. In general, the two entropies differ in one constant. What matters is that the surface of the shell collapsing coincides with area $A$ in \eqref{W0} and that it can be interpreted in the same way as $S_{BH}$. 

Thus, the entropy associated with the shell is considered as the thermodynamic entropy of equilibrium. This entropy corresponds to the information stored in the material that comes together to form a black hole, compressed in one thin layer close to the gravitational radius. Since the entropy for a given mass and area is maximized by the thermal equilibrium, we expect it to be the maximum entropy that could be stored in the material before it crosses the horizon \cite{Pretorius:1997wr}.

The fact that the entropy of a BH is also the maximum entropy that may be obtained through the Bekenstein limit was one of the main observations that led to the holographic principle \cite{Bousso:2002ju,Bekenstein:1981}. The maximum entropy stored in the shell is related to an upper bound in the information that is understood in terms of quantum bounds that are based on the principles of quantum gravity \cite{Fursaev:2004qz,Addazi:2018uhd,rovelli2017reality}.

\begin{figure}[H]
\centering
\includegraphics[width=0.5\textwidth]{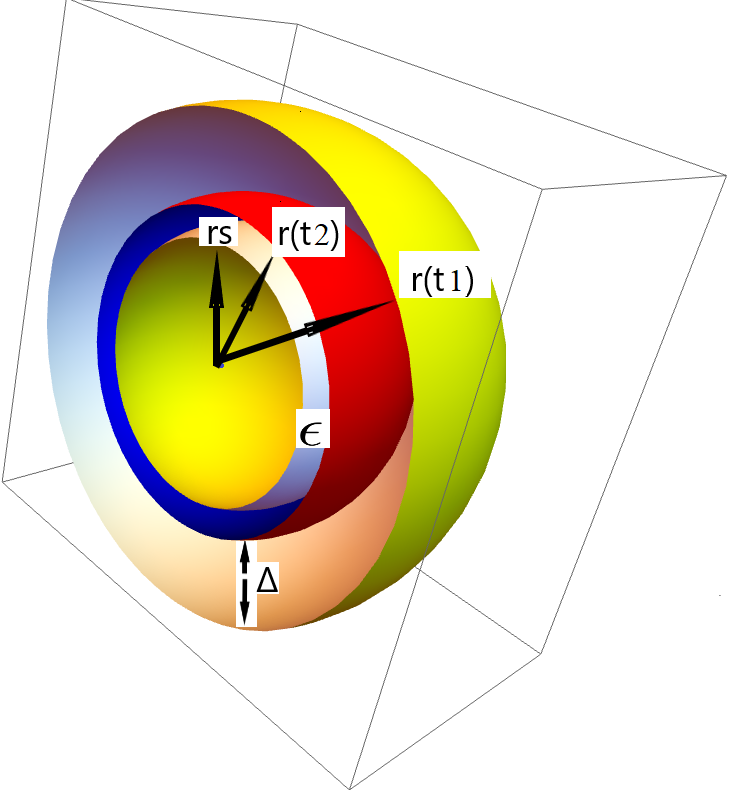}
\caption{Configuration of the vicinity of the gravitational radius.}
\label{FB300}
\end{figure}

In terms of the expression \eqref{W0}, the estimate of $\epsilon$ is introduced, which is calculated from the contribution of the area $S_{A}$ of $S_{ent}$. This contribution is determined by the quantum condition \cite{tHooft:1984kcu,Arenas:2011be}:
\begin{equation}
\epsilon\ll\Delta \ll r(t_{2}),
\label{B3}
\end{equation}
illustrated in Figure \ref{FB300}; $\Delta$ is a measure of the integration region that leads to $S_{A}$.

%%%%%%%%%%%%%%%%%%%%%%%%%%%%%%%%%%%%%%%%%%%%%%%%%%%%%%%%%%%%%%%%figura%%%%%%%%%%%%
%
%%%%%%%%%%%%%%%%%%%%%%%%%%%%%%%%%%%%%%%%%%%%%%%%%%%%%%%%%%%%%%%

Calculating $S_{ent}$, for simplicity, is done based on the scalar field in the vicinities of the shell when it is very close to its gravitational radius, resorting to quantum field theory in a curved spacetime~\cite{Mukohyama:1998rf,Arenas:2011be}. In this article, to perform this estimate, we use the particle description of quantum fields. 

%------------------------------------------------------------------------------------------------------------------------------------------
\section{Kinetic Origin of the Cutoff Parameter for a Black Shell}\label{Sec3}

The equation of movement of the shell with mass $M$ in gravitation contraction, described in Section \ref{Sec2}, is expressed by \cite{PhysRev.153.1388}:
\begin{equation}
\frac{dR}{d\tau}=\sqrt{\left[a+\frac{M}{2aR}\right]^{2}-1}
\label{B4}
\end{equation}
where $a$ is the quotient between $M$ and its mass at rest $m$, while $\tau$ is proper time in the frame of a Freely Falling Observer (FFO).

The FIDO observer that measures the collapse time of the shell is in an asymptotically flat spacetime region around the black shell:
\begin{equation}
ds^{2}=-f(r)dt^{2}+\frac{1}{f(r)}dr^{2}+r^{2}d\theta^{2}+r^{2}\sin^{2} \theta d\phi^{2},\,\,\, f(r)=1-\frac{2GM}{c^{2}r}.
\label{B0}
\end{equation}

A simple solution to the equation of movement of the shell \eqref{B4} can be expressed by the function~\cite{S.:2016dul,Akhmedov:2015xwa,Israel:1966rt}:
\begin{equation}
r(t)=r_{Sch}+\delta r e^{-t/ \bar{\tau}},
\label{B1}
\end{equation}
where $\delta r=r_{0}-r_{Sch}$, $\bar{\tau}=\frac{4GM}{3 c^{3}}$, and $r_{Sch}=\frac{2GM}{c^{2}}$.

For an external observer, the kinematics of the shell is characterized by two clearly distinguishable phases: one of rapid contraction, where the shell is far away from the gravitational radius, $r(t_{0})\gg r_{Sch}$, and a second quasi-stationary phase in the region where $r(t) \sim r_{Sch}$, and the shell's mass is concentrated around the associated horizon.

The shell's movement in the first phase of collapsing is classically described based on the General Theory of Relativity (GTR). The collapsing phase close to the gravitational radius is adjusted by introducing quantum effects within the framework of the quantum field theory in curved spacetime, such that:
\begin{equation}
\lim_{t\longrightarrow t_{2}}r(t)=r_{Sch}+\epsilon.
\label{BB1}
\end{equation}

\begin{figure}[H]
\centering
\includegraphics[width=0.5\textwidth]{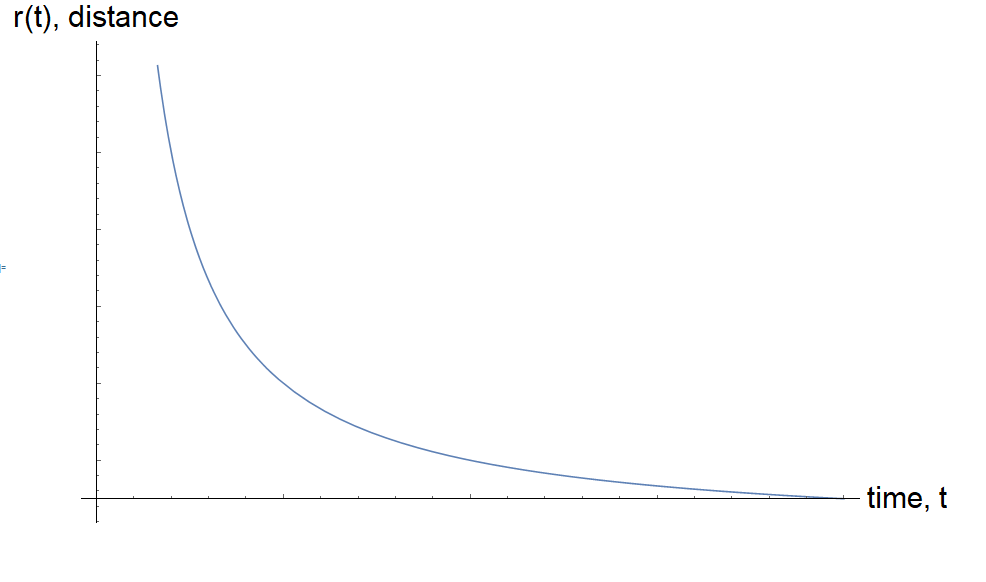}
\caption{Phase of rapid collapse.}
\label{FB400}
\end{figure}

\begin{figure}[H]
\centering
\includegraphics[width=0.5\textwidth]{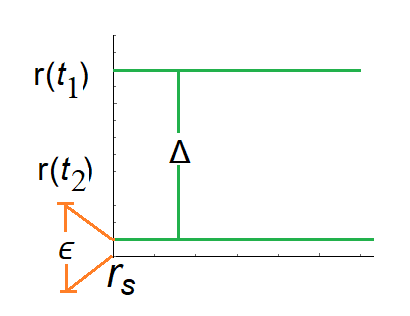}
\caption{Classical description of the gravitational collapse with a quantum bound.}
\label{FB500}
\end{figure}

The collapsing phase in the vicinity of the horizon is related to the classic phase according to Figure \ref{FB400}, where the magnification of the vertical segment on the axis $r(t)$, $\overline{r_{Sch}r(t_{1})}$, is shown in Figure~\ref{FB500}.

%%%%%%%%%%%%%%%%%%%%%%%%%%%%%%%%%%%%%%%%%%%%%%%%%%%%%%%%%%%%%%%%figura%%%%%%%%%%%%
\unskip
%
%%%%%%%%%%%%%%%%%%%%%%%%%%%%%%%%%%%%%%%%%%%%%%%%%%%%%%%%%%%%%%%%%%%%%%%%%%%%%%%%%%%

In the gravitational background given by \eqref{B0}, Expression \eqref{W0} is calculated in terms of the entropy:
\begin{equation}
S=\int_{V} sdV,
\label{B90} 
\end{equation}
where $s$ is the entropy density of the thermal atmosphere in the vicinity of the shell, which is expressed as a function of the energy density $\rho$ and pressure $P$ as:
\begin{equation}
s=\beta k_{B}\left[\rho +P\right],\,\,\,\beta=\frac{1}{k_{B}T}
\label{W10}
\end{equation}
where $k_{B}$ is the Boltzmann constant and $T$ is the local temperature given by: 
\begin{equation}
T(r)=\frac{T_{\infty}}{\sqrt{f(r(t))}}
\label{W15}
\end{equation}
for a uniform temperature $T_{\infty}$ in the asymptotically flat region.

In addition, the differential volume $dV$ corresponds to: 
\begin{equation}
dV=\frac{4\pi}{\sqrt{f(r(t))}}r^{2}(t)dr(t).
\label{B110}
\end{equation}

Entropy \eqref{B90} has two contributions: one dependent on volume $S_{V}$ for $r\gg r_{Sch}$ and another one dependent on the area $S_{A}$ for $r \sim r_{Sch}$:
\begin{equation}
S=S_{V}+S_{A}.
\label{W20}
\end{equation}

The entropy density $s$ for the contribution $S_{A}$ is calculated based on \eqref{W10} resorting to the ultrarelativistic approaches corresponding to high local temperatures, where:
\begin{equation}
P=\frac{8\pi}{h^{3}c^{3}}T^{4}k^{4}_{B},\,\,\, \rho=\frac{4\pi}{h^{3}c^{3}}T^{4}k^{4}_{B}
\label{W30}
\end{equation}
and with which we obtain \cite{Mukohyama:1998rf,Arenas:2011be,Rojas:2011ee,RojasC:2017udo}:
\begin{equation}
S=\frac{16\pi}{3h^{3}c^{2}T}\int^{\infty}_{0}\frac{E^{3}}{e^{E/k_{B}T}-\sigma}dE,
\label{W40}
\end{equation}
where $E$ is energy, $c$ is the speed of light, and $h$ is Planck's constant, with $\sigma= +1$ for bosons and $\sigma=-1$ for fermions. In the model proposed, the several values of $\sigma$ are those considering the helicities of particles in nature, which are estimated to be more than $300$ quantum fields $N$ \cite{Mukohyama:1998rf,Fursaev:2004qz}. The~study of the entropy of a black hole when considering the helicities of particles leads to the problem of species, which is still an open problem and is not clearly understood \cite{Mukohyama:1998rf,Fursaev:2004qz,Chen:2016vpx}. However, there are partial studies of entropy for scalar fields ($s=0$) \cite{Arenas:2011be,RojasC.:2018tph,Kolekar:2010py,RojasC:2017udo} and fermionic fields (s=1/2) \cite{Rojas2018Te} that confirm the dependence of entropy on the horizon area. 
\begin{equation}
 S_{BH}|_{\sigma=0, N=1} =S_{BH}|_{\sigma=1, N=1}=S_{BH}|_{\sigma=-1, N=1}=\frac{1}{4l^{2}_{p}}A. 
\end{equation}

This leads us to think that the entropy of a black hole is independent of the helicities of particles~\cite{Wald:1995yp}:
\begin{equation}
 S_{BH}|_{N=1,2,3...n}=\frac{1}{4l^{2}_{p}}A.
\end{equation}

From a mathematical point of view, the adoption of $\sigma=0$ corresponds to the argument given by Mukohyama--Israel; for the entropy density, it corresponds to the fact that the integral \eqref{W40} \cite{Mukohyama:1998rf}: ``The~purely numerical integral has the value $3!$ multiplied by $1$, $\frac{\pi^{4}}{90}$ and $\frac{7\pi^{4}}{8*90}$ for $0$, $1$ and $- 1$ respectively''. We take $\sigma=0$.

From the expressions \eqref{B90} and \eqref{W40}, we obtain:
\begin{equation}
S_{A}=\frac{16\pi}{3h^{3}c^{2}}\int_{V}dV\frac{1}{T}\int^{\infty}_{0}\frac{E^{3}}{e^{E/k_{B}T}}dE.
\label{W41}
\end{equation}
Then, the magnitude of the volume element, according to \eqref{B1}, can be written as:
\begin{equation}
\left|dV\right|=\frac{4\pi}{\sqrt{f(r(t))}}\frac{r^{2}_{s}\delta r e^{-t/\bar{\tau}}}{\bar{\tau}}dt,
\label{B140}
\end{equation}
where the small summands are disregarded in the expression:
\begin{equation}
\left|dV\right|=\frac{4\pi}{\sqrt{f(r(t))}}\left[\frac{r^{2}_{s}\delta r e^{-t/\bar{\tau}}}{\bar{\tau}}+\frac{2r_{Sch}\delta r^{2} e^{-2t/\bar{\tau}}}{\bar{\tau}}+\frac{\delta r^{3} e^{-3t/\bar{\tau}}}{\bar{\tau}}\right]dt,
\label{B120}
\end{equation}
since $ e^{-t/\bar{\tau}} \gg e^{-2t/\bar{\tau}}\gg e^{-3t/\bar{\tau}}$. \eqref{B120}

Substituting \eqref{W40} and \eqref{B140} into \eqref{B90},
\begin{equation}
S_{A}=\frac{16\pi}{3h^{3}c^{3}}\int\frac{4\pi}{\sqrt{f(r(t))}}\frac{r^{2}_{s}\delta r e^{-t/\bar{\tau}}}{\bar{\tau}}dt \frac{1}{\tau}\int^{\infty}_{0}\frac{E^{3}}{e^{E/k_{B}T}}dE
\end{equation}

Finally, based on the local temperature \eqref{W15}, $S_{A}$ is expressed as: 
\begin{equation}
S_{A}=\frac{32k_{B}^{4}c}{h^{3}}\left(\frac{T^{3}_{\infty}}{\kappa^{2}_{0}}\right)\frac{A_{H} {C_{1}}}{4\delta r},
\label{B220}
\end{equation}
where $A=4\pi r_{Sch}^{2}$, and we introduce the definition of:
\begin{equation}
 {C_{1}}=\frac{1}{\bar{ \tau}}\int^{ t_{ 2}}_{t_{1}}e^{t/\bar{\tau}} dt.
\label{B230}
\end{equation}
corresponding to the collapse time between $r(t_{1})$ and $r(t_{2})$ in the configuration of Figure \ref{FB300}. 

The expression \eqref{B220} corresponds to the known expression \eqref{W0}, where $t_{1}$ determines the finite nature of entropy in temporal terms, in contrast to $\epsilon$, which in \eqref{W0}, does the same thing in spatial terms. 

From \eqref{B1} and \eqref{B230}, we obtain a relationship between the times measured by the FIDO observer and the closeness parameter $\epsilon$ as modeled in Section \ref{Sec4}. 

%------------------------------------------------------------------------------------------------------------------------------------------
\section{Quantum Model of the Closeness Parameter}\label{Sec4}

From the expressions \eqref{B1} and \eqref{B230}:
\begin{equation}
-\int^{r(t_{2})}_{r(t_{1})}\frac{\delta r}{(r-r_{Sch})^{2}}dr=\frac{1}{\bar{\tau}}\int^{t_{2}}_{t_{1}}e^{t/\bar{\tau}}dt,
\label{B350}
\end{equation}
where $r(t_{1})=r(t_{2}) + \Delta$ and $r(t_{2}) = r_{Sch}+\epsilon$.

By integrating \eqref{B350}, we obtain:
	\[\delta r\left[ \frac{1}{\epsilon}-\frac{1}{\epsilon+\Delta}\right]=e^{t_{2}/\bar{\tau}}-e^{t_{1}/\bar{\tau}}.
\]
According to \eqref{B3}, $\epsilon\ll \Delta$. Then,
	\[\delta r\left[ \frac{1}{\epsilon}-\frac{1}{\Delta}\right]=e^{t_{2}/\bar{\tau}}-e^{t_{1}/\bar{\tau}}.
\]
	\[\delta r \left[\frac{\Delta -\epsilon}{\Delta \epsilon}\right]\sim \left[\frac{\Delta}{\epsilon \Delta}\right]\delta r=e^{t_{2}/\bar{\tau}}-e^{t_{1}/\bar{\tau}}
\]
\begin{equation}
\epsilon=\frac{\delta r}{e^{t_{2}/\bar{\tau}}-e^{t_{1}/\bar{\tau}}}.
\label{B370}
\end{equation}

The classical expression \eqref{B1} can be quantum-adjusted considering the following:

We expect that \eqref{B1}, at the limit when $t\longrightarrow t_{2}$, according to the criterion \eqref{B3}, is like the limit described by \eqref{BB1}. In other words, according to \eqref{BB1} and \eqref{B370}:
\begin{equation}
r(t_{2})=r_{Sch}+\frac{\delta r}{e^{t_{2}/\bar{\tau}}-e^{t_{1}/\bar{\tau}}},
\label{B400}
\end{equation}
which can be taken as a starting expression to generalize Expression \eqref{B1}, setting the value for $t_{1}$ and leaving $t_{2}$ as a variable:
\begin{equation}
r(t)=r_{Sch}+\frac{\delta r}{e^{t/\bar{\tau}}-e^{t_{1}/\bar{\tau}}},
\label{B400a}
\end{equation}
for which, it is required that t be finite and $t>t_{1}$, corresponding to the time of collapse measured by the FIDO.

From \eqref{B400a}, it is possible to obtain a semi-classical limit if $t_{2}\gg t_{1}$:
\begin{equation}
\epsilon=\epsilon_{Clas}=\delta re^{-t_{2}/\bar{\tau}},
\label{B410}
\end{equation}
where for the distant FIDO observer, we expect that:
	\[\epsilon \ll \kappa_{0}l^{2}_{P}.
	\]
	
The latter expression has a quantum origin due to the relationship that exists between $t$ and $\epsilon$ given by \eqref{B350}, provided $t_{2}$ is finite. A totally classic result occurs when $t_{2}$ tends continuously to infinity and, consequently, $\epsilon$ tends to zero.
For \eqref{B410}, we have $\bar{\tau}=\frac{4GM}{3c^{3}}$, and $l_{p}=\sqrt{\frac{\hbar G}{c^{3}}}$ is the Planck length. Therefore, we can write the cutoff parameter $\epsilon_{ECO}$ as:
\begin{equation}
\epsilon_{ECO}=\delta r \exp\left[\frac{-3\hbar t_{u}}{4l^{2}_{p}M}\right].
\label{B290b}
\end{equation}
Defining $x=\frac{t}{M}$, the Schwarzschild radius reduces to:
\begin{equation}
r_{Sch}=\frac{2Gxt}{c^{2}}
\end{equation}
and the initial position of the black shell becomes $r_{0} = nr_{Sch}$, where $n$ is $n=1,2,\ldots$. 

Then, the cutoff parameter is:
\begin{equation}
\epsilon=\frac{2Gxt}{c^{2}}\left[n-1\right] \exp\left[-\frac{3\hbar x}{4l^{2}_{p}}\right].
\label{B305}
\end{equation}
Therefore, we have:
\begin{equation}
 \lim_{x\rightarrow 0}\epsilon=0,
\end{equation}
For massive ECOs, the cutoff parameter value $\epsilon_{ECO}$ approaches zero.

$t_{1}$ can be estimated considering that $r(t_{2})\ll r_{0}$ and that $r(t_{2})= r_{Sch}+\epsilon\sim r_{Sch}$.

From \eqref{B1} and: 
	\[\Delta =r\left(t_{1}\right) -r\left(t_{2}\right)
\]
so:
	\[\frac{\Delta}{\delta r}=e^{-t_{1}/\bar{\tau}}-e^{-t_{2}/\bar{\tau}}
\]
	\begin{equation}
	\frac{\Delta}{r_{0}-r_{Sch}}\sim\frac{r^{2}_{s}}{r^{2}_{0}}=e^{-t_{1}/\bar{\tau}}-e^{-t_{2}/\bar{\tau}},
	\label{B415}
\end{equation}
where: 
	\begin{equation}
	\frac{\Delta}{r(t_{2})}\sim\frac{r(t_{2})}{r_{0}},\,\,\, \frac{\Delta}{r_{Sch}}\sim\frac{r_{Sch}}{r_{0}}\Rightarrow \Delta \sim \frac{r_{Sch}^{2}}{r_{0}}.
	\label{B420}
\end{equation}
\eqref{B370} can be written as:
\begin{equation}
\epsilon=\frac{\delta r}{De^{(t_{1}+t_{2})/\bar{\tau}}},
\label{B430}
\end{equation}
with $D=\frac{r^{2}_{s}}{r^{2}_{0}}$.

At the limit $t_{1}\ll t_{2}$, \eqref{B430} must coincide with \eqref{B410}. Comparing these last two expressions, we obtain:
\begin{equation}
e^{t_{1}/\bar{\tau}}=\frac{r^{2}_{0}}{r^{2}_{s}}
\label{B440}
\end{equation}

Then, from \eqref{B370} and \eqref{B440}:
\begin{equation}
\epsilon=\frac{\delta r}{e^{t_{2}/\bar{\tau}}-\frac{r^{2}_{0}}{r^{2}_{s}}}.
\label{B450}
\end{equation}

In \eqref{B450}, we included a preliminary criterion given by \eqref{B420} that can be adjusted observationally in such a way that $\epsilon$ expressed by \eqref{B450}, which is clearly greater than its corresponding magnitude given by \eqref{B410}, approaches $\kappa_{0}l^{2}_{p}$.

Comparing \eqref{B410} and \eqref{B450}, we obtain:
\begin{equation}
\frac{\epsilon_{Clas}}{\epsilon}=1-\frac{r^{2}_{0}}{r^{2}_{s}}e^{-t_{2}/\bar{\tau}}; 
 \label{B460} 
\end{equation}
thus, for \eqref{B460}, it is possible to define $P=1-\frac{r^{2}_{0}}{r^{2}_{s}}e^{-t_{2}/\bar{\tau}}$. Therefore, \eqref{B460} is reduced to:
\begin{equation}
\frac{\epsilon_{Clas}}{\epsilon}=P. 
 \label{B470} 
\end{equation}
$P$ is subject to the following conditions:

\begin{itemize}
		\item If $P=P_{Max}=1$, $\epsilon_{Clas}=\epsilon$; therefore, the quantum effects are not relevant in estimating the cutoff.
 \item 	If $P\longrightarrow 0$, the quantum effects are very significant in estimating the cutoff.
\end{itemize}

%------------------------------------------------------------------------------------------------------------------------------------------
\section{Summary and Discussion}\label{Sec5}

ECOs have become very important objects of research on the nature and existence of astrophysical BHs, considering that these objects are not ruled out by the recent gravitational wave observations~\cite{Abbott:2020khf}. In addition, the closeness parameter $\epsilon$ that characterizes the ECOs would be associated with any basic scale of quantum gravity \cite{Addazi:2018uhd}. In that context, this paper contributes to the conceptualization of the origin of the cutoff parameter $\epsilon$, with a model that characterizes the ECOs according to the frame of the observer \cite{Oppenheimer:1939}.

The model proposed consists of a thin spherical dust shell that gravitationally contracts from a specific position $r_{0}$ to close to its gravitational radius $r(t_{2})=r_{Sch}+\epsilon$, at a finite time $t_{2}$ measured in the frame of an FIDO.

The cutoff makes two types of contributions: a classical one governed by General Relativity and~one of a quantum nature, if the ECO is very close to the horizon, when estimating that the maximum entropy is contained within the material that composes the shell. Such entropy coincides with the Bekenstein--Hawking entropy. The established cutoff corresponds to a dynamic quantity dependent on the coordinate time that is measured by a Fiducial Observer (FIDO).

Based on the bound that determines the maximum information that the shell can store compressed in a thin layer close to the gravitational radius, we estimate $\epsilon\ll \kappa_{0}l^{2}_{p}$ in a semi-classical approach and a limit of $\epsilon$ tending to $\kappa_{0}l^{2}_{p}$ when the quantum effects are significant. 

The maximum information associated with the shell is quantified with the contribution of area $S_{A}$ to its entropy \cite{Pretorius:1997wr}. With this expression, we introduce a suggestive relationship between the microscopic world and the macroscopic world. 

Usually, $S_{A}$ is calculated in a non-divergent manner in terms of a cutoff $\epsilon$, as shown in Expression~\eqref{W0}, and it is adjusted with a quantum gravity criterion to calculate the Bekenstein--Hawking entropy $S_{BH}$. In this article, we propose a divergence control in terms of a finite time $t_{2}$, according to the definition of $ {C_{1}}$ in \eqref{B230}. What is interesting is that the two criteria are related through the expression \eqref{B350} and, more specifically, based on \eqref{B450}. In other words, since the entropy $S_{BH}$ is proportional to the shell's entropy at the limit when $r(t)\sim r_{Sch}$, astrophysical measurements of $t_{2}$ and $r_{0}$ would allow estimating the cutoff $\epsilon$. With this relationship, we have an observational criterion to compare the assumptions made based on the principles of quantum gravity. 

On the other hand, the estimates of $\epsilon$ can be compared with those respectively calculated in the articles \cite{Barausse:2018vdb,Guo:2017jmi,Harada:2018zfg} and also related to the parameters of the different types of ECOs through the relationship of $\epsilon$ with the Love numbers \cite{Addazi:2018uhd}.

The ECO model being proposed depends on external observation and is compatible with BHs in two senses: with respect to the FIDOs, this type of ECO is undistinguishable from a BH, and regarding an FFO observer, the shell collapses into a BH. In this latter case, the explanation of $S_{BH}$ requires a complete theory of quantum gravity. In the case of the description in terms of the external observers at rest with respect to the associated horizon, without knowing the details about quantum gravity, it is possible to interpret $S_{BH}$ as a thermodynamic entropy.

BH formation, maturity, and death are still not completely understood; total understanding may establish the existence of a quantum gravity theory. Specifically, a BH is characterized by a singularity surrounded by an event horizon, which hides the interior of the BH from Fiducial Observers (FIDOs)~\cite{Penrose:1969pc,Thorne1986}. This is key because it distinguishes a BH from other types of ECOs in the frame of an FIDO. This implies that the geometry outside the ECO is consistent with that of a BH \cite{Cardoso:2019rvt}.

It is noteworthy to mention that the role of the FIDO is remarkable in this study since the cutoff parameter is only measured by such observers \cite{Cardoso:2019rvt,Addazi:2018uhd}.

This model can be improved if a dynamic spacetime, such as a Kerr spacetime, is considered because the cutoff parameter in Equation \eqref{B1} takes into account observational data \cite{Ferrara:2014wua,2019ApJ...879L...3B}.

%%%%%%%%%%%%%%%%%%%%%%%%%%%%%%%%%%%%%%%%%%%%%%%%%%%%%%%%%%%%%%%%%%%%%
%%%%%%%%%%%%%%%%%%%%%%%%%%%%%%%%%%%%%%%%%%%%%%%%
\section{Acknowledgments}
This work was supported by  the Departamento de Administrativo de Ciencia, Tecnolog\'ia e Innovaci\'on, Colciencias.
%%%%%%%%%%%%%%%%%%%%%%%%%%%%%%%%%%%%%%%%%%%%%%%%%%%%%%%%%%%%%%%%%%%%%
%%%%%%%%%%%%%%%%%%%%%%%%%%%%%%%%%%%%%%%%%%%%%%%%%%%%%%%%%%%%%%%


\begin{thebibliography}{29}

\bibitem[Penrose(1969)]{Penrose:1969pc}
Penrose, R. Gravitational collapse: The role of general relativity.
\newblock \emph{NCimR} \textbf{1969}, \emph{1}, 252.

\bibitem{Penrose:1964wq}
Penrose, R. Gravitational collapse and spacetime singularities.
\emph{Phys. Rev. Lett.} \textbf{1965}, \emph{14}, 57--59.

\bibitem{Cardoso:2019rvt}
{Cardoso, V.; Pani, P. Testing the nature of dark compact objects: A status report.
\newblock {\em Living Rev. Rel.} \textbf{2019}, \emph{22}, 1--104.}



\bibitem{Chapline:2000en}
Chapline, G.; Hohlfeld, E.; Laughlin, R.B.; Santiago, D.I. Quantum phase transitions and the breakdown of classical general relativity.
\emph{Int. J. Mod. Phys. A} \textbf{2003}, \emph{18}, 3587--3590.


\bibitem{S.:2016dul}
Robel Arenas, S.J.; Castro, O.F. Euclidean Approach for Entropy of Black Shells. \emph{arXiv} \textbf{2016}, arXiv:1606.06786.


\bibitem{Poisson:2009pwt}
Poisson, E.
\emph{A Relativist's Toolkit: The Mathematics of Black-Hole Mechanics}; Cambridge University Press: Cambridge, UK, 2004.


\bibitem{Abbott:2016}
Abbott, B.P.; Abbott, R.; Abbott, T.D.; Abernathy, M.R.; Acernese, F.; Ackley, K.; Adams, C.; Adams, T.; Addesso, P.; Adhikari, R.X.; et al. Observation of Gravitational Waves from a Binary Black Hole Merger.
\newblock {\em Phys. Rev. Lett.} \textbf{2016}, \emph{116}, 061102. 

\bibitem{Abbott:2020uma}
Abbott, B.P.; Abbott, R.; Abbott, T.D.; Abraham, S.; Acernese, F.; Ackley, K.; Adams, C.; Adhikari, R.X.; Adya,~V.B.; Affeldt, C.; et al. GW190425: Observation of a Compact Binary Coalescence with Total Mass $\sim3.4 M_{\odot}$.
\newblock \emph{Astrophys. J. Lett.} \textbf{2020}, \emph{892}, L3.

\bibitem{Abbott:2017}
Abbott, B.P.; Abbott, R.; Abbott, T.D.; Acernese, F.; Ackley, K.; Adams, C.; Adams, T.; Addesso, P.; Adhikari,~R.X.; Adya, V.B.; et al. GW170817: Observation of Gravitational Waves from a Binary Neutron Star Inspiral.
\emph{Phys. Rev. Lett.} \textbf{2017}, \emph{119}, 161101.



\bibitem{Abbott:2020khf}
Abbott, R.; Abbott, T.D.; Abraham, S.; Acernese, F.; Ackley, K.; Adams, C.; Adhikari, R.X.; Adya, V.B.; Affeldt,~C.; Agathos, M.; et al. GW190814: Gravitational Waves from the Coalescence of a 23 Solar Mass Black Hole with a 2.6 Solar Mass Compact Object.
\newblock \emph{Astrophys. J.} \textbf{2020}, \emph{896}, L44.

\bibitem{Abedi:2016hgu}
Abedi, J.; Dykaar, H.; Afshordi, N. Echoes from the abyss: Tentative evidence for Planck-scale structure at black hole horizons.
\newblock \emph{Phys. Rev.} \textbf{2017}, \emph{D9}6, 082004.

\bibitem{Cardoso:2016oxy}
Cardoso, V.; Hopper, S.; Macedo, C.F.; Palenzuela, C.; Pani, P. Gravitational-wave signatures of exotic compact objects and of quantum corrections at the horizon scale.
\newblock { \em Phys. Rev. D} \textbf{2016}, \emph{94}, 084031.

\bibitem{Addazi:2018uhd}
Addazi, A.; Marcianò, A.; Yunes, N. Can we probe Planckian corrections at the horizon scale with gravitational waves?
 \newblock {\em Phys. Rev. Lett.} \textbf{2019}, \emph{122}, 081301.


\bibitem{Barausse:2018vdb}
Barausse, E.; Brito, R.; Cardoso, V.; Dvorkin, I.; Pani, P. The stochastic gravitational-wave background in the absence of horizons.
\newblock \emph{Class. Quant. Grav.} \textbf{2018}, \emph{35}, 20LT01.




\bibitem{Chen:2019hfg}
Chen, B.; Chen, Y.; Ma, Y.; Lo, K.L.R.; Sun, L. Instability of Exotic Compact Objects and Its Implications for Gravitational-Wave Echoes. \emph{arXiv} \textbf{2019}, arXiv:1902.08180.

\bibitem{Pani:2015hfa}
Pani, P.; Gualtieri, L.; Maselli, A.; Ferrari, V. Tidal deformations of a spinning compact object.
\newblock { \em Phys. Rev. D} \textbf{2015}, \emph{92}, 024010.

\bibitem{Binnington:2009bb}
Binnington, T.; Poisson, E. Relativistic theory of tidal Love numbers.
\newblock {\em Phys. Rev. D} \textbf{2009}, \emph{80}, 084018.

\bibitem{Pani:2015nua}
Pani, P.; Gualtieri, L.; Ferrari, V. Tidal Love numbers of a slowly spinning neutron star.
\newblock { \em Phys. Rev. D} \textbf{2015}, \emph{92}, 124003.

\bibitem{Cardoso:2017cfl}
Cardoso, V.; Franzin, E.; Maselli, A.; Pani, P.; Raposo, G. Testing strong-field gravity with tidal Love numbers.
\newblock { \em Phys. Rev. D} \textbf{2017}, \emph{95}, 084014.

\bibitem{Yagi:2016ejg}
Yagi, K.; Yunes, N. I-Love-Q Relations: From Compact Stars to Black Holes.
\newblock { \em Class. Quant. Grav.} \textbf{2016}, \emph{33}, 095005.

\bibitem{LeTiec:2020spy}
Le Tiec, A.; Casals, M. Spinning Black Holes Fall in Love. \emph{arXiv} \textbf{2020}, arXiv:2007.00214.

\bibitem{Guo:2017jmi}
Guo, B.; Hampton, S.; Mathur, S.D. Can we observe fuzzballs or firewalls?
\newblock { \em JHEP} \textbf{2018}, \emph{7}, 162.

\bibitem{Harada:2018zfg}
Harada, T.; Cardoso, V.; Miyata, D. Particle creation in gravitational collapse to a horizonless compact object.
\newblock { \em Phys. Rev. D} \textbf{2019}, \emph{99}, 044039.

\bibitem{Addazi:2019bjz}
  {Addazi, A.; Marcianò, A.; Yunes, N. Gravitational Instability of Exotic Compact Objects.
\newblock { \em Eur. Phys. J. C} \textbf{2020}, \emph{80}, 36.
}
\bibitem{Paranjape:2009ib}
Paranjape, A.; Padmanabhan, T. Radiation from collapsing shells, semiclassical backreaction and black hole formation.
\newblock { \em Phys. Rev. D} \textbf{2009}, \emph{80}, 044011.


\bibitem{Saravani:2012is}
Saravani, M.; Afshordi, N.; Mann, R.B. Empty black holes, firewalls, and the origin of Bekenstein\textendash{}Hawking entropy.
\newblock { \em Int. J. Mod. Phys. D} \textbf{2015}, \emph{23}, 1443007.

\bibitem{Barcelo:2010xk}
Barceló; C.; Liberati, S.; Sonego, S.; Visser, M. Hawking-like radiation from evolving black holes and compact horizonless objects.
\newblock { \em JHEP} \textbf{2011}, \emph{2011}, 1--30.

\bibitem{Uchikata:2015yma}
Uchikata, N.; Yoshida, S. Slowly rotating thin shell gravastars.
\newblock { \em Class. Quant. Grav.} \textbf{2016}, \emph{33}, 025005.

\bibitem{Cardoso:2008kj}
Cardoso, V.; Pani, P.; Cadoni, M.; Cavaglia, M. Instability of hyper-compact Kerr-like objects.
\newblock {\em Class. Quant. Grav.} \textbf{2008}, \emph{25}, 195010.

\bibitem{Cardoso:2007az}
Cardoso, V.; Pani, P.; Cadoni, M.; Cavaglia, M. Ergoregion instability of ultracompact astrophysical objects.
\newblock {\em Phys. Rev. D} \textbf{2008}, \emph{77}, 124044.


\bibitem{Cardoso:2014sna}
Cardoso, V.; Crispino, L.C.; Macedo, C.F.; Okawa, H.; Pani, P. Light rings as observational evidence for event horizons: Long-lived modes, ergoregions and nonlinear instabilities of ultracompact objects.
\newblock { \em Phys. Rev. D} \textbf{2014}, \emph{90}, 044069.




\bibitem{Fursaev:2004qz}
 Fursaev, D.V. Can one understand black hole entropy without knowing much about quantum gravity?
\newblock {\em Phys. Part. Nucl.} \textbf{2005}, \emph{36}, 81.


\bibitem{Mukohyama:1998rf}
Mukohyama, S.; Israel, W. Black holes, brick walls and the Boulware state.
\newblock {\em Phys. Rev.} \textbf{1998}, \emph{D58}, 104005.
\bibitem{tHooft:1984kcu}
 Hooft, G.T. On the Quantum Structure of a Black Hole.
\newblock {\em Nucl. Phys. B} \textbf{1985}, \emph{256}, 727.
\bibitem{Israel:1976}
Israel, W. Thermo field dynamics of black holes.
\newblock {\em Phys. Lett. A} \textbf{1976}, \emph{57}, 107. 

%42
\bibitem{Zurek:1985}
Zurek, W.H.; Thorne, K.S. Statistical Mechanical Origin of the Entropy of a Rotating, Charged Black Hole.
\newblock {\em Phys. Rev. Lett.} \textbf{1985}, \emph{54}, 2171.
%43
\bibitem{Fulling:1987}
Fulling, S.A.; Ruijsenaars, S.N.M. Temperature, periodicity and horizons.
\newblock {\em Phys. Rep. } \textbf{1987}, \emph{152}, 135.
%44
	\bibitem{Arenas:2011be}
 {Arenas S, J. R. and Tejeiro J.~M.~}
 Entanglement Entropy of Black Shells.
\newblock \emph{Nuovo Cim.} \textbf{2010}, \emph{B125}, 1223.
%45
\bibitem{Pretorius:1997wr}
 {Pretorius, F.; Vollick, D.; Israel, W. An Operational approach to black hole entropy.
\newblock {\em Phys. Rev.} \textbf{1998}, \emph{D57}, 6311.}

%46
\bibitem{Bousso:2002ju}
Bousso, R. The Holographic principle.
\newblock {\em Rev. Mod. Phys.} \textbf{2002}, \emph{74}, 825.
%47
\bibitem{Bekenstein:1981}
 Bekenstein, J.D. Universal upper bound on the entropy-to-energy ratio for bounded systems.
\newblock {\em Phys. Rev. D} \textbf{1981}, \emph{23}, 287.
%48
\bibitem{rovelli2017reality}
Rovelli, C.
\newblock \emph{Reality is Not What it Seems: The Journey to Quantum Gravity};
\newblock Penguin Books: London, UK, 2017; 
\newblock ISBN 9780735213920.


%49
\bibitem{PhysRev.153.1388}
Israel, W. Gravitational Collapse and Causality.
\newblock \emph{Phys. Rev.} \textbf{1967}, \emph{153}, 1388.



%50
\bibitem{Akhmedov:2015xwa}
Akhmedov, E.T.; Godazgar, H.; Popov, F.K. Hawking radiation and secularly growing loop corrections.
\newblock \emph{Phys. Rev. D} \textbf{2016}, \emph{93}, 024029.

%51
\bibitem{Israel:1966rt}
Israel, W. Singular hypersurfaces and thin shells in general relativity.
\newblock \emph{Nuovo Cim. B} \textbf{1966}, \emph{44}, 1--14.
%52
\bibitem{Rojas:2011ee}
 {Rojas C, W. A. and Arenas S, J. R.} C{\'o}mo se afecta la descripci{\'o}n termodin{\'a}mica de los sistemas f{\'i}sicos cuando se incluye la gravedad?  arXiv:1110.4058.
%53
\bibitem{RojasC:2017udo}
 {Rojas C, W. A. and Arenas S, J. R. Black Shells, Dirac's Field and the species problem.} , arXiv:1712.08724. %ref 47 and ref 51 are the same please confirm
\bibitem{Chen:2016vpx}
Chen, Y.Z.; Li, W.D.; Dai, W.S. Why the entropy of spacetime is independent of species of particles--the species problem?
\emph{Eur. Phys. J. C} \textbf{2018}, \emph{78}, 635.
%35
\bibitem{RojasC.:2018tph}
Rojas, W.A.; Arenas, J. R. Thermodynamics of Hot Quantum Scalar Field in a $(D + 1)$ Dimensional Curved. Spacetime. \emph{Electron. J. Theor. Phys.} \textbf{2018}, \emph{14}, 115--124.


%36
\bibitem{Kolekar:2010py}
Kolekar, S.; Padmanabhan, T. Ideal Gas in a strong Gravitational field: Area dependence of Entropy.
\emph{Phys. Rev. D} \textbf{2011}, \emph{83}, 064034.



\bibitem{Rojas2018Te}
Rojas Castillo, W.A. 
Mec{\'a}nica Estad{\'i}stica de la Termodin{\' a}mica de Black Shells. Ph.D. Thesis,  {Departamento de F\'isica, Universidad  Nacional  de Colombia } Available online: 
\url{https://repositorio.unal.edu.co/handle/unal/63982}  {(November 30, 2020 ).} %Please add accessed date and the location

\bibitem{Wald:1995yp}
 Wald, R.M. \emph{Quantum Field Theory in Curved Space-Time and Black Hole Thermodynamics};  {University of Chicago Press: Chicago, IL, USA, 1994.} %Newly added information, please confirm.
 \bibitem{Oppenheimer:1939}
Oppenheimer, J. R.; Snyder, H. On Continued Gravitational Contraction.
\newblock \emph{Phys. Rev.} \textbf{1939}, \emph{56}, 455 .



%55
\bibitem{Thorne1986}
Thorne, K.S.; Thorne, K.S.; Price, R.H.; MacDonald, D.A.
 \newblock \emph{Black Holes: The Membrane Paradigm}; Yale University Press: London, UK, 1986.
	
%56	
	\bibitem{Ferrara:2014wua}
Ferrara, A.; Salvadori, S.; Yue, B.; Schleicher, D. Initial mass function of intermediate mass black hole seeds.
\newblock \emph{Mon. Not. Roy. Astron. Soc.} \textbf{2014}, \emph{443}, 2410.

%57
	\bibitem{2019ApJ...879L...3B}
Basu, S.; Das, A. The Mass Function of Supermassive Black Holes in the Direct-collapse Scenario.
\newblock \emph{Astrophy. J.} \textbf{2019}, \emph{879}, L3.

\end{thebibliography}
\end{document}